\documentclass[aps,prx,superscriptaddress,twocolumn,floatfix]{revtex4-1}
\usepackage{graphicx}
\usepackage{epsfig}
\usepackage{amsmath}
\usepackage{amssymb}
\usepackage{times}

\begin{document}

\title{Ultrafast Calculation of Diffuse Scattering from Atomistic Models}

\author{Joseph A. M. Paddison}
\email[]{jamp3@cam.ac.uk}

\affiliation{Churchill College, University of Cambridge, Storey's Way, Cambridge CB3 0DS, United Kingdom}

\date{\today}

\begin{abstract}
Diffuse scattering is a rich source of information about disorder
in crystalline materials, which can be modelled using atomistic techniques
such as Monte Carlo and molecular dynamics simulations. Modern X-ray
and neutron scattering instruments can rapidly measure large volumes
of diffuse-scattering data. Unfortunately, current algorithms for
atomistic diffuse-scattering calculations are too slow to model large
data sets completely, because the fast Fourier transform (FFT) algorithm
has long been considered unsuitable for such calculations \cite{Butler_1992}.
Here, I present a new approach for ultrafast calculation of atomistic
diffuse-scattering patterns. I show that the FFT can be actually be
used to perform such calculations rapidly, and that a fast method
based on sampling theory can be used to reduce high-frequency noise
in the calculations. I benchmark these algorithms using realistic
examples of compositional, magnetic, and displacive disorder. They
accelerate the calculations by a factor of at least 100,
making refinement of atomistic models to large diffuse-scattering
volumes practical.
\end{abstract}

\maketitle

\section{Introduction\label{sec:introduction}}

Disorder plays an increasingly important role in our understanding
of crystalline materials. Whereas conventional crystallography is
primarily concerned with the average positions of atoms or molecules,
local deviations from the average structure are fundamental to the
properties of many important systems. Topical examples include fast-ion
conductors \cite{Keen_2002}, frustrated magnets \cite{Fennell_2009,Paddison_2013a},
the polar nanodomains of lead-based perovskite ferroelectrics \cite{Xu_2004,Pasciak_2012},
and the orbital correlations of colossal magnetoresistance manganites
\cite{Shimomura_1999}. Diffuse scattering\textemdash the weak features
observed beneath and between the Bragg peaks in scattering experiments\textemdash plays
a central role in helping us to understand structural disorder \cite{Billinge_2007}.
Whereas Bragg peaks arise from the ideal periodicity of the average
structure and yield information about single-particle correlations,
diffuse scattering arises from local deviations from the average structure
and yields information about pairwise correlations. In disordered
crystals, strong pairwise correlations are often present at the nanoscale,
yielding highly-structured diffuse-scattering patterns \cite{Keen_2015,Overy_2016,Weber_2012}.
Disorder in crystalline materials can often be divided into three
broad categories: local atomic displacements, which may be either
static or dynamic; local variations in chemical composition
such as site mixing and atomic vacancies; and disordered
spin arrangements in magnetic materials. In all cases, the modulation
of the diffuse-scattering intensity provides vital information about
how the relevant degrees of freedom\textemdash whether displacive,
compositional, or magnetic\textemdash are locally correlated.

Recent advances in instrumentation have allowed rapid collection of
large three-dimensional (3D) diffuse-scattering datasets from single
crystals. In neutron-scattering experiments, such datasets can be
measured in a matter of days using time-of-flight instruments with
large detector coverage, such as SXD at the ISIS Neutron Source \cite{Keen_2006}
and Corelli at the Oak Ridge National Laboratory \cite{Ye_2018}. In X-ray scattering experiments, single-photon-counting
area detectors such as the Pilatus \cite{Kraft_2009} allow such datasets
to be obtained in minutes. These experimental developments have been
coupled with advances in 3D atomistic modelling. Atomistic models
may be generated using Monte Carlo or molecular dynamics simulations
based on a set of interaction parameters, which are determined either
from first-principles simulations \cite{Gutmann_2013} or by fitting
to experimental data \cite{Welberry_1998,Welberry_2008,Weber_2002}.
Alternatively, atomistic models may be generated by fitting atomic
positions or magnetic-moment orientations directly to experimental
data, in an approach called reverse Monte Carlo refinement \cite{McGreevy_1988,Tucker_2007,Paddison_2012}.
Improvements in computing power have greatly reduced the time required
to generate atomistic models using these approaches.

A key step in atomistic diffuse-scattering analysis is calculating
the diffuse-scattering intensity, which must usually be repeated several
times as the model is improved. In principle, this involves the straightforward
procedure of taking the Fourier transform of a set of atomic positions.
In practice, however, such calculations are limitingly slow, because
of two main problems. The first problem is that the fast Fourier transform
(FFT) algorithm\textemdash which can yield enormous increases in speed
compared to direct Fourier summation\textemdash has been considered
unsuitable for these calculations \cite{Butler_1992}. This is because
the FFT requires sampling atomic positions on an equally-spaced grid
\cite{Cooley_1965}, but a grid that sampled local atomic displacements
accurately would need so many samples that the computational advantages
of the FFT would be negated \cite{Butler_1992,Neder_1989,Rahman_1989}.
The second problem is that diffuse-scattering calculations are often
marred by high-frequency noise. This occurs because the entire atomistic
model is assumed to scatter coherently, which often allows the many
essentially-uncorrelated atomic displacements at large interatomic
distances to dominate the calculation \cite{Butler_1992}. Current
software addresses this problem by dividing the supercell into many
smaller regions called ``sub-boxes'' or ``lots'' and averaging
the scattering intensity over sub-boxes \cite{Butler_1992,Proffen_1997}.
This approach is effective at reducing noise, but has a large computational
cost that can only be reduced by parallelization \cite{Gutmann_2010}.
Because of their computational cost, diffuse-scattering calculations
have typically been restricted to a few 2D sections of reciprocal
space \cite{Welberry_2008}. In most cases, therefore, the information
content of 3D diffuse-scattering data has yet to be fully explored.

Here, I present an approach to diffuse-scattering calculations that
seeks to address the computational limitations of current algorithms.
I present two main results. First, I critically reassess the assumption
that the FFT is not useful for atomistic diffuse-scattering calculations.
I show that, on the contrary, the FFT can be used to perform such
calculations rapidly. This FFT-based approach is exact for systems in which
the disorder is compositional, magnetic, or in which atomic displacements
are drawn from a discrete set of values; if atomic displacements can
take a continuous range of values, then the FFT can be applied to
a specified order of approximation. Second, I show that the desirable
noise-reduction properties of the ``sub-box'' approach can also
be obtained by a faster method based on sampling theory. I demonstrate
the practicality of these algorithms by presenting example calculations
based on real materials, and by providing a program \textsc{Scatty} 
for ultrafast diffuse-scattering calculations [27]. 
These developments
accelerate atomistic diffuse-scattering calculations by several orders
of magnitude compared to current algorithms, making refinement of
atomistic models to 3D diffuse-scattering data a practical possibility.

This article is structured as follows. In Section \ref{sec:atomistic},
I review the scattering equations and aspects of sampling theory in
the context of atomistic models. In Section \ref{sec:fft}, I show
how the FFT can be used to accelerate atomistic diffuse-scattering
calculations. In Section \ref{sec:examples}, I present example calculations
for model systems that exhibit occupational, magnetic, and displacive
disorder, in which highly-structured diffuse scattering is driven
by ``ice rules''. In Section \ref{sec:resampling}, I show how noise
in the calculations can be reduced using a resampling approach. I
conclude in Section \ref{sec:conclusions} with a discussion of the
implications of this work.

\section{Atomistic simulations\label{sec:atomistic}}

In atomistic simulations, disorder is modelled using a ``virtual
crystal''. This is a supercell of the crystallographic unit cell,
typically containing $N\sim10^{4}$ atomic positions, in which each
atomic position is decorated by variables corresponding to its occupancy
by a certain element, its displacement from its average position,
and/or the orientation of its magnetic moment. I consider a supercell
that consists of $n_{1}n_{2}n_{3}$ crystallographic unit cells, where
$n_{1}$, $n_{2}$, and $n_{3}$ are the numbers of unit cells parallel
to the $\mathbf{a},\mathbf{b}$, and $\mathbf{c}$ crystal axes, respectively.
An atomic position in the supercell is given by
\begin{equation}
\mathbf{r}=n_{1}r_{1}\mathbf{a}+n_{2}r_{2}\mathbf{b}+n_{3}r_{3}\mathbf{c};\quad0\leq r_{\alpha}<1.\label{eq:r_def}
\end{equation}
An interatomic vector is denoted $\Delta\mathbf{r}$, with components
$\Delta r_{\alpha}$. A general wavevector is given by
\begin{align}
\mathbf{Q} & =h\mathbf{a}^{*}+k\mathbf{b}^{*}+l\mathbf{c}^{*}\label{eq:}\\
 & \equiv\frac{Q_{1}}{n_{1}}\mathbf{a}^{*}+\frac{Q_{2}}{n_{2}}\mathbf{b}^{*}+\frac{Q_{3}}{n_{3}}\mathbf{c}^{*};\quad Q_{\alpha}\in\mathbb{R},\label{eq:Q_def}
\end{align}
where reciprocal-lattice vectors are defined as $\mathbf{a}^{*}=2\pi\mathbf{b}\times\mathbf{c}/V$,
etc., where $V$ is the unit-cell volume. Atomistic simulations usually
impose periodic boundary conditions to avoid edge effects, and I assume
throughout this article that this is the case. The Bragg positions
of the periodic supercell are then given by
\begin{align}
\mathbf{G} & =\frac{G_{1}}{n_{1}}\mathbf{a}^{*}+\frac{G_{2}}{n_{2}}\mathbf{b}^{*}+\frac{G_{3}}{n_{3}}\mathbf{c}^{*};\quad G_{\alpha}\in\mathbb{Z},\label{eq:G_def}
\end{align}
and become increasingly closely spaced as the number of unit cells
in the supercell is increased.

The coherent neutron-scattering intensity from a
real crystal is given, in the kinematic approximation, by
\begin{align}
I(\mathbf{Q}) & =\frac{1}{N}\sum_{j,k}\left\langle b_{j}b_{k}\exp\left[\mathrm{i}\mathbf{Q}\cdot(\mathbf{r}_{j}-\mathbf{r}_{k})\right]\right\rangle \label{eq:intensity_dble_sum}\\
 & \equiv\sum_{\Delta\mathbf{r}}\rho(\Delta\mathbf{r})\exp(\mathrm{i}\mathbf{Q}\cdot\Delta\mathbf{r}),
\end{align}
where $b_{j}$ is the coherent neutron scattering length of atom $j$,
and $\rho(\Delta\mathbf{r})=\frac{1}{N}\sum_{j,k}\left\langle b_{j}b_{k}\delta_{\Delta\mathbf{r},\mathbf{r}_{j}-\mathbf{r}_{k}}\right\rangle $
is the 3D pair-distribution function \cite{Weber_2012}. Angle brackets
denote spatial and temporal averaging, which can be approximated in
atomistic modelling by averaging over many supercells. For X-ray scattering,
the $b_{j}$ are replaced by $|\mathbf{Q}|$-dependent X-ray form
factors, $f_{j}(|\mathbf{Q}|)$. Conventional diffraction measurements do not resolve the energy of the scattered beam and therefore integrate over energy; this integration is complete provided that the incident radiation energy significantly exceeds the energy of structural fluctuations in the sample. The $\rho(\Delta\mathbf{r})$ obtained from energy-integrated data is the correlation function of the instantaneous (equal-time) atomic positions \cite{Egami_2003}, and atomistic models of energy-integrated data similarly function as ``snapshots" of the instantaneous atomic positions \cite{Goodwin_2004}.

By factorizing the double summation
in Eq.~(\ref{eq:intensity_dble_sum}), the scattering intensity
can be rewritten as
\begin{equation}
I(\mathbf{Q})=\frac{1}{N}\left\langle |F(\mathbf{Q})|^{2}\right\rangle ,\label{eq:intensity_sf}
\end{equation}
where the structure factor is a discrete Fourier transform,
\begin{equation}
F(\mathbf{Q})=\sum_{j=1}^{N}b_{j}\exp(\mathrm{i}\mathbf{Q}\cdot\mathbf{r}_{j}).\label{eq:sf}
\end{equation}
In general, the scattering intensity from a disordered crystal can
be separated into its Bragg and diffuse contributions, 
\begin{align}
I(\mathbf{Q}) & =I_{\mathrm{Bragg}}(\mathbf{Q})+I_{\mathrm{diffuse}}(\mathbf{Q}),\label{eq:bragg_diffuse}\\
 & =\frac{1}{N}\left|\left\langle F(\mathbf{Q})\right\rangle \right|^{2}+\frac{1}{N}\left\langle \left|F(\mathbf{Q})-\left\langle F(\mathbf{Q})\right\rangle \right|^{2}\right\rangle ,\label{eq:bragg_diffuse_sf}
\end{align}
which arise from the average structure and local modulations away
from the average, respectively \cite{Frey_1997}.

Atomistic models inevitably contain far fewer unit cells ($\sim$$10^{3}$)
than do real crystals ($\sim$$10^{23}$), and it is important to
consider how this affects scattering calculations. Fundamentally,
a periodic supercell only contains information about pair correlations
for which $-r_{\mathrm{cut}}\leq\Delta r_{\alpha}<r_{\mathrm{cut}}$,
where $r_{\mathrm{cut}}=0.5$. Scattering calculations may account
for this in two different ways. First, if the scattering is calculated
using Eq.~(\ref{eq:intensity_dble_sum}), then the separation
between pairs is taken to be the shortest separation between their
periodic images; this is called the nearest-image convention and involves
replacing $\Delta r_{\alpha}$ with $\Delta r_{\alpha}\pm1$ where
necessary to ensure $-r_{\mathrm{cut}}\leq\Delta r_{\alpha}<r_{\mathrm{cut}}$.
This procedure can be conceptualized as multiplying the pair-distribution
function for an infinite tiling of the supercell, $\rho_{\mathrm{inf}}(\Delta r_{\alpha})$,
by a cutoff function, so that
\begin{equation}
\rho(\Delta r_{\alpha})=\rho_{\mathrm{inf}}(\Delta r_{\alpha})\times\mathrm{cutoff}(\Delta r_{\alpha}),\label{eq:pdf_cutoff}
\end{equation}
where $\mathrm{cutoff}(\Delta r_{\alpha})$ is a rectangular function
equal to $1$ for $-r_{\mathrm{cut}}\leq\Delta r_{\alpha}<r_{\mathrm{cut}}$
and zero elsewhere. In this case, the calculated $I(\mathbf{Q})$
is a smooth function of wavevector, which contains finite-size artifacts
because the contribution of the average structure to $\rho_{\mathrm{inf}}(\Delta r_{\alpha})$
is truncated. In contrast, $I_{\mathrm{diffuse}}(\mathbf{Q})$ is
free from finite-size artifacts provided that the local modulations
are short-ranged compared to $r_{\mathrm{cut}}$. The second approach
is to calculate the scattering using Eqs.~(\ref{eq:intensity_sf})
and (\ref{eq:sf}). In this case, the nearest-image convention cannot
be applied. Consequently, the scattering intensity may only be sampled
at the supercell Bragg positions $\{\mathbf{G}\}$, for which $I(\mathbf{G})$
is unchanged by the nearest-image convention because $\exp(\pm\mathrm{i}\mathbf{G}\cdot\mathbf{n})=1$.
The two approaches become equivalent for real crystals, which are
large enough that the boundary conditions become irrelevant.

In practice, it is much faster to calculate the scattering intensity
using Eqs.~(\ref{eq:intensity_sf}) and (\ref{eq:sf})\textemdash which
require only a single summation over atomic positions\textemdash than
to use Eq.~(\ref{eq:intensity_dble_sum}), which requires a
double summation. However, the resulting limitation that only $I(\mathbf{G})$
may be calculated is often too restrictive, especially when a direct
comparison with experimental data is required \cite{Butler_1992}.
To solve this problem, I consider the information content of the scattering
pattern. Applying the convolution theorem to Eq.~(\ref{eq:pdf_cutoff})
yields 
\begin{align}
I(\mathbf{Q}) & =\sum_{\mathbf{G}=-\infty}^{\infty}I(\mathbf{G})W_{\mathrm{sinc}}(\mathbf{Q}-\mathbf{G}),\label{eq:intensity_convol}
\end{align}
where the weight function
\begin{align}
W_{\mathrm{sinc}}(\mathbf{Q}-\mathbf{G}) & =8r_{\mathrm{cut}}^{3}\prod_{\alpha=1}^{3}\textrm{sinc}\left[2\pi r_{\mathrm{cut}}(Q_{\alpha}-G_{\alpha})\right]\label{eq:weight_sinc}
\end{align}
is the Fourier transform of the rectangular function, $I(\mathbf{G})$
is the Fourier transform of $\rho_{\mathrm{inf}}(\Delta\mathbf{r})$
sampled at $\{\mathbf{G}\}$, and $\mathrm{sinc}(x)=\sin(x)/x$ \cite{Whittaker_1915,Shannon_1949}.
In the context of sampling theory, Eqs.~(\ref{eq:intensity_convol})
and (\ref{eq:weight_sinc}) are known as the Whittaker-Shannon interpolation
formulae. They show that the scattering intensity can, in principle,
be reconstructed at any wavevector $\mathbf{Q}$, given only its samples
at the supercell Bragg positions $\{\mathbf{G}\}$. This is possible
because the supercell Bragg positions sample $I(\mathbf{Q})$ at its
Nyquist rate \cite{Sayre_1952}. 

In the rest of this article, I will first show how $I(\mathbf{G})$
can be calculated rapidly using the FFT algorithm. I will then show
how $I(\mathbf{Q})$ can be estimated in practice, by modifying Eqs.~(\ref{eq:intensity_convol}) and (\ref{eq:weight_sinc}).

\section{Fast Fourier transform \label{sec:fft}}

\subsection{Requirements for the FFT}

The FFT is an algorithm that calculates the discrete Fourier transform
rapidly \cite{Cooley_1965,Singleton_1969}. In the 3D case relevant
for crystalline materials, it calculates a function of the form
\begin{align}
X_{\mathbf{k}} & =\sum_{\mathbf{R}}x_{\mathbf{R}}\exp(\mathrm{i}\mathbf{k}\cdot\mathbf{R}).\label{eq:fft}
\end{align}
Unlike the discrete Fourier transform, for which the position vector
$\mathbf{R}$ and wavevector \textbf{$\mathbf{k}$} can take any real
values, the FFT imposes two restrictions. First, the $x_{\mathbf{R}}$
must be arranged on an equally-spaced grid, so that
\begin{equation}
\mathbf{R}=R_{1}\mathbf{a}+R_{2}\mathbf{b}+R_{3}\mathbf{c};~R_{\alpha}\in\left\{ 0,1,...,N_{\alpha}-1\right\} ,\label{eq:r_condition}
\end{equation}
where $N_{\alpha}$ denotes the number of grid points parallel to
crystal axis $\alpha$ in the supercell. Second, the values of $\mathbf{k}$
at which the Fourier coefficients $X_{\mathbf{k}}$ may be evaluated
are given by
\begin{equation}
\mathbf{k}=\frac{k_{1}}{N_{1}}\mathbf{a}^{\ast}+\frac{k_{2}}{N_{2}}\mathbf{b}^{\ast}+\frac{k_{3}}{N_{3}}\mathbf{c}^{\ast};\quad k_{\alpha}\in\left\{ 0,1,...,N_{\alpha}-1\right\} .\label{eq:k_condition}
\end{equation}
Below, I will identify the grid points with lattice points, so that
$N_{\alpha}=n_{\alpha}$ and each grid point is the origin of a particular
unit cell within the supercell. In this case, the values of $\mathbf{k}$
given by Eq.~(\ref{eq:k_condition}) are a subset of the supercell
Bragg positions that lie within a fundamental domain of the reciprocal
lattice of the supercell. 

The computational cost of the discrete Fourier transform is on the
order of $\left(N_{1}N_{2}N_{3}\right)^{2}$, whereas for the FFT
it is on the order of $N_{1}N_{2}N_{3}\log_{2}(N_{1}N_{2}N_{3})$\textemdash a
huge saving for large $N_{1}N_{2}N_{3}$. Unfortunately, the atomic
positions in a crystal do not generally satisfy Eq.~(\ref{eq:r_condition}),
except in the special case where atoms occupy a Bravais lattice without
disorder. In early work, it was assumed that this restriction could
only be addressed by using a fine grid encompassing all the atomic
positions to some specified accuracy; however, the number of grid
points required in 3D was found to be be prohibitively large \cite{Butler_1992,Neder_1989,Rahman_1989}.
Consequently, the FFT has been considered unsuitable for atomistic
diffuse-scattering calculations, and is not implemented in current
software such as \textsc{Diffuse} \cite{Butler_1992}, \textsc{Discus}
\cite{Proffen_1997}, \textsc{ZMC} \cite{Goossens_2011}, and\textsc{
Zods} \cite{Michels-Clark_2014}. 

I now show that the scattering intensity can actually be rewritten
in a form that allows the FFT to be efficiently applied. My approach makes 
use of the underlying periodicity of the average structure to obtain the FFT in time proportional to $n_{1}n_{2}n_{3}\log_{2}(n_{1}n_{2}n_{3})$, where $n_{1}n_{2}n_{3}$ is the number of unit cells, compared to  $N_{1}N_{2}N_{3}\log_{2}(N_{1}N_{2}N_{3})$ in the original FFT analysis, where $N_{1}N_{2}N_{3}$ is a very large number of grid points \cite{Butler_1992}. To emphasise
this underlying periodicity, I write the position of an atom in the supercell
as
\begin{equation}
\mathbf{r}_{\mathbf{R},\mu,i}\equiv\mathbf{R}+\mathbf{r}_{\mu}+\mathbf{u}_{\mathbf{R},\mu,i},\label{eq:r_vec}
\end{equation}
where $\mathbf{R}$ henceforth denotes a lattice point given by Eq.~(\ref{eq:r_condition})
with $N_{\alpha}=n_{\alpha}$; $\mathbf{r}_{\mu}$ is the average
position of site $\mu$ within the crystallographic unit cell; and
$\mathbf{u}_{\mathbf{R},\mu,i}$ is the local displacement of the
atom of element $i$ belonging to site $\mu$ at lattice point $\mathbf{R}$.
A Bragg position of the supercell can be written as
\begin{equation}
\mathbf{G}\equiv\mathbf{H}+\mathbf{k},\label{eq:g_vec}
\end{equation}
where $\mathbf{k}$ henceforth denotes a wavevector given by Eq.~(\ref{eq:k_condition})
with $N_{\alpha}=n_{\alpha}$, and 
\begin{equation}
\mathbf{H}=H_{1}\mathbf{a}^{\ast}+H_{2}\mathbf{b}^{\ast}+H_{3}\mathbf{c}^{\ast};\quad h_{\alpha}\in\mathbb{Z}\label{eq:h_condition}
\end{equation}
is a Bragg position of the crystallographic unit cell. By substituting
Eqs.~(\ref{eq:r_vec}) and (\ref{eq:g_vec}) into Eq.~(\ref{eq:sf}),
and using the fact that $\exp(\mathrm{i}\mathbf{H}\cdot\mathbf{R})=1$,
the structure factor can be expressed as
\begin{equation}
F(\mathbf{G})=\sum_{\mathbf{R},\mu,i}\delta_{\mathbf{R},\mu,i}b_{\mu,i}\exp[\mathrm{i}\mathbf{G}\cdot(\mathbf{r}_{\mu}+\mathbf{u}_{\mathbf{R},\mu,i})]\exp(\mathrm{i}\mathbf{k}\cdot\mathbf{R}),\label{eq:total_sf}
\end{equation}
where $\delta_{\mathbf{R},\mu,i}$ is equal to $1$ if site $\mu$
at lattice point $\mathbf{R}$ is occupied by an atom of element $i$,
and is otherwise zero, and $b_{\mu,i}$ is a coherent neutron-scattering
length. I also define the difference between the local occupancy and
the average occupancy,
\begin{align}
a_{\mathbf{R},\mu,i} & =\frac{\delta_{\mathbf{R},\mu,i}-c_{\mu,i}}{c_{\mu,i}},\label{eq:occ}
\end{align}
where $c_{\mu,i}$ is the average occupancy of site $\mu$ by atoms
of element $i$. This separation of the structure into an average
part and a local modulation allows the Bragg and diffuse contributions
to the strucure factor to be separated. From Eqs.~(\ref{eq:total_sf})
and (\ref{eq:occ}), the structure factor is given by
\begin{equation}
F(\mathbf{G})=\sum_{\mu,i}[U_{\mathbf{k},\mu,i}(\mathbf{G})+A_{\mathbf{k},\mu,i}(\mathbf{G})]c_{\mu,i}b_{\mu,i}\exp(\mathrm{i}\mathbf{G}\cdot\mathbf{r}_{\mu}),\label{eq:total_sf_fin}
\end{equation}
where I have defined a pair of Fourier transforms for each site $\mu$
and element $i$,
\begin{align}
U_{\mathbf{k},\mu,i}(\mathbf{G}) & =\sum_{\mathbf{R}}\exp(\mathrm{i}\mathbf{G}\cdot\mathbf{u}_{\mathbf{R},\mu,i})\exp(\mathrm{i}\mathbf{k}\cdot\mathbf{R}),\label{eq:U_ft}\\
A_{\mathbf{k},\mu,i}(\mathbf{G}) & =\sum_{\mathbf{R}}a_{\mathbf{R},\mu,i}\exp(\mathrm{i}\mathbf{G}\cdot\mathbf{u}_{\mathbf{R},\mu,i})\exp(\mathrm{i}\mathbf{k}\cdot\mathbf{R}),\label{eq:A_ft}
\end{align}
which are also used in the ``modulation wave'' approach to diffuse-scattering
analysis \cite{Withers_2015}. The Bragg structure factor is given
by
\begin{equation}
\left\langle F(\mathbf{G})\right\rangle =n_{1}n_{2}n_{3}\delta_{\mathbf{G},\mathbf{H}}\sum_{\mu,i}T_{\mu,i}(\mathbf{G})c_{\mu,i}b_{\mu,i}\exp(\mathrm{i}\mathbf{G}\cdot\mathbf{r}_{\mu}),\label{eq:bragg_sf}
\end{equation}
where
\begin{equation}
T_{\mu,i}(\mathbf{G})=\frac{1}{n_{1}n_{2}n_{3}}\sum_{\mathbf{R}}\left\langle(1+a_{\mathbf{R},\mu,i})\exp(\mathrm{i}\mathbf{G}\cdot\mathbf{u}_{\mathbf{R},\mu,i})\right\rangle \label{eq:dw}
\end{equation}
is the Debye-Waller factor. I will show below how the FFT can be applied
to calculate Eq.~(\ref{eq:total_sf_fin}).

\subsection{Compositional/occupational disorder}

I consider first a scenario in which the disorder is purely compositional
or occupational (I will use these terms interchangeably), so that
all atomic displacements $\mathbf{u}_{\mathbf{R},\mu,i}$ can be set
to zero. In this case, the structure factor given by Eq.~(\ref{eq:total_sf_fin})
reduces to
\begin{align}
F_{\mathrm{c}}(\mathbf{G}) & =\left\langle F_{\mathrm{c}}(\mathbf{G})\right\rangle +\sum_{\mu,i}a_{\mathbf{k},\mu,i}c_{\mu,i}b_{\mu,i}\exp(\mathrm{i}\mathbf{G}\cdot\mathbf{r}_{\mu}),\label{eq:comp_sf_diffuse}
\end{align}
in which 
\begin{equation}
a_{\mathbf{k},\mu,i}=\sum_{\mathbf{R}}a_{\mathbf{R},\mu,i}\exp(\mathrm{i}\mathbf{k}\cdot\mathbf{R})\label{eq:comp_ft}
\end{equation}
is given by Eq.~(\ref{eq:A_ft}) with $\mathbf{u}_{\mathbf{R},\mu,i}=0$,
and $\left\langle F_{\mathrm{c}}(\mathbf{G})\right\rangle $ is given
by Eq.~(\ref{eq:bragg_sf}) with $T_{\mathbf{R},\mu,i}(\mathbf{G})=1$.
The diffuse intensity from compositional disorder is then given by
\begin{equation}
I_{\mathrm{c,diffuse}}(\mathbf{G})=\frac{1}{N}\left\langle \left|F_{\mathrm{c}}(\mathbf{G})-\left\langle F_{\mathrm{c}}(\mathbf{G})\right\rangle \right|{}^{2}\right\rangle ,\label{eq:comp_diffuse_intensity}
\end{equation}
where angle brackets denote averaging over supercells. More effective
averaging is obtained by using many smaller supercells instead of
a few larger ones; I recall that the former approach does not produce
finite-size artifacts in diffuse-scattering calculations, provided
that $r_{\mathrm{cut}}$ for the supercell exceeds the correlation
length of the local modulations.

A key observation of this article is that Eq.~(\ref{eq:comp_ft})
has the same form as Eq.~(\ref{eq:fft}), with the substitutions
$x_{\mathbf{R}}=a_{\mathbf{R},\mu,i}$ and $X_{\mathbf{k}}=a_{\mathbf{k},\mu,i}$.
Consequently, Eq.~(\ref{eq:comp_ft}) can be readily calculated
by the FFT. This is possible because each site $\mu$ in the unit
cell forms a Bravais lattice; hence, Eq.~(\ref{eq:comp_ft})
is periodic in reciprocal space \cite{Borie_1971,Hayakawa_1975}.
Eq.~(\ref{eq:comp_ft}) can also be applied to models in which
local atomic displacements are present, provided they can take only
relatively few different magnitudes and directions. In such cases,
the number of atoms in the crystallographic unit cell is increased
according to the number of possible displacements, and each displaced
atomic position is assigned to a site in the crystallographic unit
cell; in this way, displacive disorder is mapped to occupational disorder
in a ``split-site'' model. Because atomistic models of real
materials often consider discrete displacement distributions as a first approximation \cite{Welberry_2008},
Eq.~(\ref{eq:comp_ft}) is often relevant in practice.

The computer time required to calculate $F(\mathbf{G})$ at $N_{\mathbf{G}}$
wavevectors using this FFT-based approach is approximately given by the sum
of two terms, a structure-factor term $t^{\prime}n_{\mathrm{uc}}N_{\mathbf{G}}$ and an FFT term $t^{\prime\prime}n_{\mathrm{uc}}n_{1}n_{2}n_{3}\log_{2}(n_{1}n_{2}n_{3})$,
which represent the
times required for calculations of Eqs.~(\ref{eq:comp_sf_diffuse})
and (\ref{eq:comp_ft}), respectively, where $n_{\mathrm{uc}}$ is the
number of atoms in the unit cell. Application of the discrete Fourier
transform to Eq.~(\ref{eq:total_sf}) requires an approximate time $t^{\prime}n_{\mathrm{uc}}n_{1}n_{2}n_{3}N_{\mathbf{G}}$.
In typical simulations, $n_{1}n_{2}n_{3}\sim10^{3}$, $n_{\mathrm{uc}}\sim10$,
and $N_{\mathbf{G}}$ may range from $\sim$$10^{3}$ for scattering
planes to more than $10^{6}$ for large volumes. In practice, I will
show in section \ref{sec:examples} that the FFT-based approach accelerates typical
scattering calculations by two to three orders of magnitude compared
to the traditional approach.

\subsection{Compositional/occupational and displacive disorder}

If atomic displacements can take many possible magnitudes or directions,
the structure factor given by Eq.~(\ref{eq:total_sf_fin}) cannot
be directly evaluated by the FFT, because $U_{\mathbf{k},\mu,i}(\mathbf{G})$
and $A_{\mathbf{k},\mu,i}(\mathbf{G})$ contain factors of $\exp(\mathrm{i}\mathbf{G}\cdot\mathbf{u}_{\mathbf{R},\mu,i})$
and hence do not have the same form as Eq.~(\ref{eq:fft}). I
now show that this problem can be addressed by expanding $\exp(\mathrm{i}\mathbf{G}\cdot\mathbf{u}_{\mathbf{R},\mu,i})$
as a Taylor series,
\begin{equation}
\exp(\mathrm{i}\mathbf{G}\cdot\mathbf{u}_{\mathbf{R},\mu,i})=\sum_{n=0}^{\infty}\frac{\mathrm{i}^{n}}{n!}(\mathbf{G}\cdot\mathbf{u}_{\mathbf{R},\mu,i})^{n}.\label{eq:disp_expansion}
\end{equation}
Writing the scalar product in terms of its components yields
\begin{align}
\mathbf{G}\cdot\mathbf{u}_{\mathbf{R},\mu,i} & =2\pi\sum_{\alpha=1}^{3}G_{\alpha}u_{\mathbf{R},\mu,i}^{\alpha},\label{eq:gdotu_1}\\
(\mathbf{G}\cdot\mathbf{u}_{\mathbf{R},\mu,i})^{2} & =(2\pi)^2\sum_{\alpha_{1},\alpha_{2}}G_{\alpha_{1}}G_{\alpha_{2}}u_{\mathbf{R},\mu,i}^{\alpha_{1}}u_{\mathbf{R},\mu,i}^{\alpha_{2}},\label{eq:gdotu_2}\\
(\mathbf{G}\cdot\mathbf{u}_{\mathbf{R},\mu,i})^{3} & =(2\pi)^3\sum_{\alpha_{1},\alpha_{2},\alpha_{3}}G_{\alpha_{1}}G_{\alpha_{2}}G_{\alpha_{3}}u_{\mathbf{R},\mu,i}^{\alpha_{1}}u_{\mathbf{R},\mu,i}^{\alpha_{2}}u_{\mathbf{R},\mu,i}^{\alpha_{3}},\label{eq:gdtou_3}
\end{align}
and so on; evidently, the term of order $n$ in the Taylor expansion
contains a sum of all products of $n$ components of $\mathbf{G}$
and $n$ components of $\mathbf{u}_{\mathbf{R},\mu,i}$. Crucially,
the products $u_{\mathbf{R},\mu,i}^{\alpha_{1}}u_{\mathbf{R},\mu,i}^{\alpha_{2}}...u_{\mathbf{R},\mu,i}^{\alpha_{n}}$
depend only on atomic position, and their Fourier transforms can therefore
be evaluated by the FFT. The generalization of Eqs.~(\ref{eq:gdotu_1})\textendash (\ref{eq:gdtou_3})
for the term of order $n$ is
\begin{align}
(\mathbf{G}\cdot\mathbf{u}_{\mathbf{R},\mu,i})^{n} & =\sum_{\boldsymbol{\alpha}(n)}\mathcal{G}_{\boldsymbol{\alpha}(n)}\mathcal{U}_{\mathbf{R},\mu,i}^{\boldsymbol{\alpha}(n)},\label{eq:gu_expansion}
\end{align}
where $\mathcal{G}_{\boldsymbol{\alpha}(n)}=(2\pi)^n\prod_{j=1}^{n}G_{\alpha_{j}}$, $\mathcal{U}_{\mathbf{R},\mu,i}^{\boldsymbol{\alpha}(n)}=\prod_{j=1}^{n}u_{\mathbf{R},\mu,i}^{\alpha_{j}}$,
and $\mathcal{G}_{\boldsymbol{\alpha}(0)}=\mathcal{U}_{\boldsymbol{\alpha}(0)}\equiv1$.
Substituting Eqs.~(\ref{eq:disp_expansion}) and (\ref{eq:gu_expansion})
into Eqs.~(\ref{eq:U_ft}) and (\ref{eq:A_ft}), and the results
into Eq.~(\ref{eq:total_sf_fin}), I obtain the structure factor
\begin{multline}
F(\mathbf{G})=\sum_{\mu,i}c_{\mu,i}b_{\mu,i}\exp(\mathrm{i}\mathbf{G}\cdot\mathbf{r}_{\mu}) \\
\times \sum_{n=0}^{\infty}\sum_{\boldsymbol{\alpha}(n)}\frac{\mathrm{i}^{n}}{n!}(\mathcal{U}_{\mathbf{k},\mu,i}^{\boldsymbol{\alpha}(n)}+\mathcal{A}_{\mathbf{k},\mu,i}^{\boldsymbol{\alpha}(n)})\mathcal{G}_{\boldsymbol{\alpha}(n)},\label{eq:total_sf_taylor}
\end{multline}
where I have defined the Fourier transforms
\begin{align}
\mathcal{U}_{\mathbf{k},\mu,i}^{\boldsymbol{\alpha}(n)} & =\sum_{\mathbf{R}}\mathcal{U}_{\mathbf{R},\mu,i}^{\boldsymbol{\alpha}(n)}\exp(\mathrm{i}\mathbf{k}\cdot\mathbf{R}),\label{eq:U_fft}\\
\mathcal{A}_{\mathbf{k},\mu,i}^{\boldsymbol{\alpha}(n)} & =\sum_{\mathbf{R}}a_{\mathbf{R},\mu,i}\mathcal{U}_{\mathbf{R},\mu,i}^{\boldsymbol{\alpha}(n)}\exp(\mathrm{i}\mathbf{k}\cdot\mathbf{R}).\label{eq:A_fft}
\end{align}
Eqs.~(\ref{eq:U_fft}) and (\ref{eq:A_fft}) have the same form
as Eq.~(\ref{eq:fft}) and can therefore be evaluated by the
FFT. The Bragg structure factor is given by
\begin{multline}
\left\langle F(\mathbf{G})\right\rangle =\delta_{\mathbf{G},\mathbf{H}}\sum_{\mu,i}c_{\mu,i}b_{\mu,i}\exp(\mathrm{i}\mathbf{G}\cdot\mathbf{r}_{\mu})\\
\times \sum_{n=0,2}^{\infty}\sum_{\boldsymbol{\alpha}(n)}\frac{\mathrm{i}^{n}}{n!}\left\langle \mathcal{U}_{\mathbf{k},\mu,i}^{\boldsymbol{\alpha}(n)}+ \mathcal{A}_{\mathbf{k},\mu,i}^{\boldsymbol{\alpha}(n)}\right\rangle \mathcal{G}_{\boldsymbol{\alpha}(n)}
,\label{eq:bragg_sf_disp}
\end{multline}
where only even-$n$ terms are included in the Taylor expansion because
the odd-$n$ terms average to zero. Finally, the diffuse intensity
is obtained as
\begin{equation}
I_{\mathrm{diffuse}}(\mathbf{G})=\frac{1}{N}\left\langle \left|F(\mathbf{G})-\left\langle F(\mathbf{G})\right\rangle \right|{}^{2}\right\rangle .\label{eq:disp_diffuse_intensity}
\end{equation}

In practice, it is necessary to truncate the expansion of $\exp(\mathrm{i}\mathbf{G}\cdot\mathbf{u})$
at a finite order of approximation. An upper bound on the magnitude of the remainder for a Taylor expansion of $\exp(\mathrm{i}\mathbf{G}\cdot\mathbf{u})$ to order $n$ is given by 
$(\mathbf{G}\cdot\mathbf{u})^{n+1}/(n+1)!$, allowing the worst-case error in the approximation to be estimated for a given (e.g., maximal) value of $\mathbf{G}\cdot\mathbf{u}$. The number of FFTs that must be
performed to evaluate the term of order $n$ is equal to $(n+1)(n+2)/2$,
which increases rapidly with $n$. Nevertheless, because of the favourable
scaling of the FFT, calculations with negligible loss of accuracy
still allow large performance improvements over the discrete Fourier
transform, as I will show in section \ref{sec:examples}.

\subsection{Magnetic disorder}

I now consider a magnetic system in which magnetic moments $\mathbf{M}{}_{\mathbf{R},\mu,i}$
decorate atomic positions. Because the neutron has a magnetic moment,
neutron scattering is a powerful technique to study local magnetic
correlations in materials such as spin liquids, spin glasses, and
frustrated magnets. The magnetic structure factor for neutron scattering
is a vector quantity \cite{Squires_1978,Lovesey_1987},
\begin{equation}
\mathbf{F}_{\mathrm{mag}}(\mathbf{G})=\sum_{\mu,i}f_{\mu,i}^{\mathrm{mag}}(|\mathbf{G}|)T_{\mu,i}(\mathbf{G})\mathbf{M}_{\mathbf{k},\mu,i}\exp(\mathrm{i}\mathbf{G}\cdot\mathbf{r}_{\mu}),\label{eq:mag_sf}
\end{equation}
where $f_{\mu,i}^{\mathrm{mag}}(|\mathbf{G}|)$ is the magnetic form
factor for atoms of element $i$ belonging to site $\mu$, and coupling
between magnetic and displacive variables is included only via the
Debye-Waller factor $T_{\mu,i}(\mathbf{G})$. The Fourier transform
of the magnetic moments is given by
\begin{equation}
\mathbf{M}_{\mathbf{k},\mu,i}=\sum_{\mathbf{R}}\mathbf{M}_{\mathbf{R},\mu,i}\exp(\mathrm{i}\mathbf{k}\cdot\mathbf{R}).\label{eq:mag_ft}
\end{equation}
It is straightforward to calculate Eq.~(\ref{eq:mag_ft}) using
the FFT, as has been noted previously \cite{Conlon_2010}; the only
difference in implementation between compositional and magnetic cases
is that the FFT is applied for each component of the vector $\mathbf{M}_{\mathbf{R},\mu,i}$
in the magnetic case. The magnetic neutron-scattering intensity is
given by
\begin{equation}
I_{\mathrm{mag}}(\mathbf{G})=\frac{C}{N}\left\langle \left|\mathbf{F}_{\mathrm{mag}}^{\perp}(\mathbf{G})\right|^{2}\right\rangle ,\label{eq:mag_intensity}
\end{equation}
where the constant $C=(\gamma_{\mathrm{n}}r_{0}/2)^{2}=0.07265$\,barn,
and
\begin{equation}
\mathbf{F}_{\mathrm{mag}}^{\perp}=\hat{\mathbf{G}}\times\mathbf{F}_{\mathrm{mag}}\times\hat{\mathbf{G}}\label{eq:mag_projection}
\end{equation}
is the projection of the magnetic structure factor perpendicular to
$\hat{\mathbf{G}}=\mathbf{G}/|\mathbf{G}|$. If the magnetic moments
do not show long-range order, then the magnetic scattering will be
entirely diffuse; this is the case in, e.g., spin-ice materials \cite{Bramwell_2001,Fennell_2009}. 

\section{Example calculations\label{sec:examples}}

To benchmark the FFT-based approach, I present diffuse-scattering
calculations for simple examples of occupational, magnetic, and displacive
disorder. In all cases, I use the same underlying model of ``ice
rules'' on a pyrochlore network of corner-sharing tetrahedra.

As an example of occupational disorder, I consider an idealized model
of proton disorder in cubic water ice $I_{\mathrm{c}}$ \cite{Pauling_1935}.
Oxygen atoms are ordered and lie at the centres of the tetrahedra.
Protons are disordered according to the ``ice rule'' that each oxygen
has two covalently-bonded protons close to it, such that these two
O---H bonds point towards vertices of the tetrahedron. The
model is disordered because there are six ways of satisfying this
constraint on each tetrahedron, which gives rise to the well-known
zero-point entropy of water ice \cite{Pauling_1935,Giauque_1936}.
The disorder is occupational because tetrahedra have four possible
proton sites, of which only two are occupied for any given tetrahedron.

As an example of magnetic disorder, I consider the spin-ice state
observed in materials such as Ho$_{2}$Ti$_{2}$O$_{7}$ and Dy$_{2}$Ti$_{2}$O$_{7}$,
in which magnetic Ho$^{3+}$ or Dy$^{3+}$ ions occupy a pyrochlore
network. The Ho$^{3+}$/Dy$^{3+}$ magnetic moments are constrained
by the crystalline electric field to point towards or away from the
tetrahedra centres, and the magnetic interactions are such that, at
low temperatures, two moments on each tetrahedron point towards its
centre and two point away \cite{Bramwell_2001}. The resulting degeneracy
of spin arrangements leads to a zero-point entropy equivalent to that
of water ice \cite{Ramirez_1999}.

Finally, as an example of displacive disorder, I consider the pyrochlore
material Y$_{2}$Mo$_{2}$O$_{7}$ \cite{Gardner_1999}, in which
an ``orbital ice'' state was recently proposed. In the proposed
model, Jahn-Teller-active Mo$^{4+}$ ions locally displace parallel
to Mo\textemdash Mo vectors such that two Mo$^{4+}$ ions on each
tetrahedron displace towards each other \cite{Thygesen_2017}. The
resulting degeneracy is again equivalent to water ice and spin ice.
The ice-rules models are all characterized by the selection of a single
edge on each tetrahedron, which may be identified with the two protons
close to the oxygen in water ice, the two spins pointing away from
the tetrahedron centre in water ice, or the two Mo$^{4+}$ ions that
displace towards each other in Y$_{2}$Mo$_{2}$O$_{7}$ \cite{Keen_2015}.

I simulated the diffuse scattering from each model as follows. First,
I generated $100$ cubic supercells that obeyed the ice rule using
a Metropolis Monte Carlo algorithm. Each supercell contained $1000$
conventional unit cells ($16\thinspace000$ vertices of the pyrochlore
network). Second, I mapped these supercells to the disorder models
discussed above. For the water-ice model, I used a lattice constant
$a=6.35$\,\AA, a covalent O\textemdash H bond length of $0.95$\,\AA,
and the neutron-scattering length for $^{2}$H. For the spin-ice model,
I used $a=10.0$\,\AA~and the neutron magnetic form factor for Ho$^{3+}$.
For the orbital-ice model, I used $a=10.0$\,\AA, a Mo displacement
magnitude of $0.1$\,\AA~\cite{Thygesen_2017}, and the atomic X-ray
form factor for Mo. 
Finally, I calculated the diffuse scattering in the $(hhl)$ plane on a grid of $401\times401$ pixels, which extended over the range $-6\leq h\leq6$ and $6\sqrt{2}\leq l\leq6\sqrt{2}$ and contained $\sim$$10^{4}$ supercell Bragg positions. 
All calculations averaged the scattering intensities over the 100 supercells to reduce statistical noise. I implemented the FFT using an updated version of the Fortran code
by Singleton \cite{Singleton_1969}. For calculations based on the FFT, I used Eqs.~(\ref{eq:comp_sf_diffuse})\textendash(\ref{eq:comp_diffuse_intensity})
for the water-ice model, and Eqs.~(\ref{eq:mag_sf})\textendash (\ref{eq:mag_projection})
for the spin-ice model. For the orbital-ice model, I calculated the diffuse intensity in two different ways using the FFT: first, by mapping
the displacive disorder to occupational disorder and applying the
exact Eqs.~(\ref{eq:comp_sf_diffuse})\textendash(\ref{eq:comp_diffuse_intensity});
and second, by applying a fifth-order Taylor approximation using Eqs.~(\ref{eq:total_sf_taylor})\textendash(\ref{eq:disp_diffuse_intensity}).
For comparison, I also used the discrete Fourier transform to calculate
Eqs.~(\ref{eq:total_sf}) and (\ref{eq:mag_sf}) without taking
advantage of reciprocal-space periodicity. I performed all the calculations
on a laptop with a $2.9$\,GHz Intel Core i5 processor, and measured the CPU
time. 

\begin{figure*} 
\includegraphics{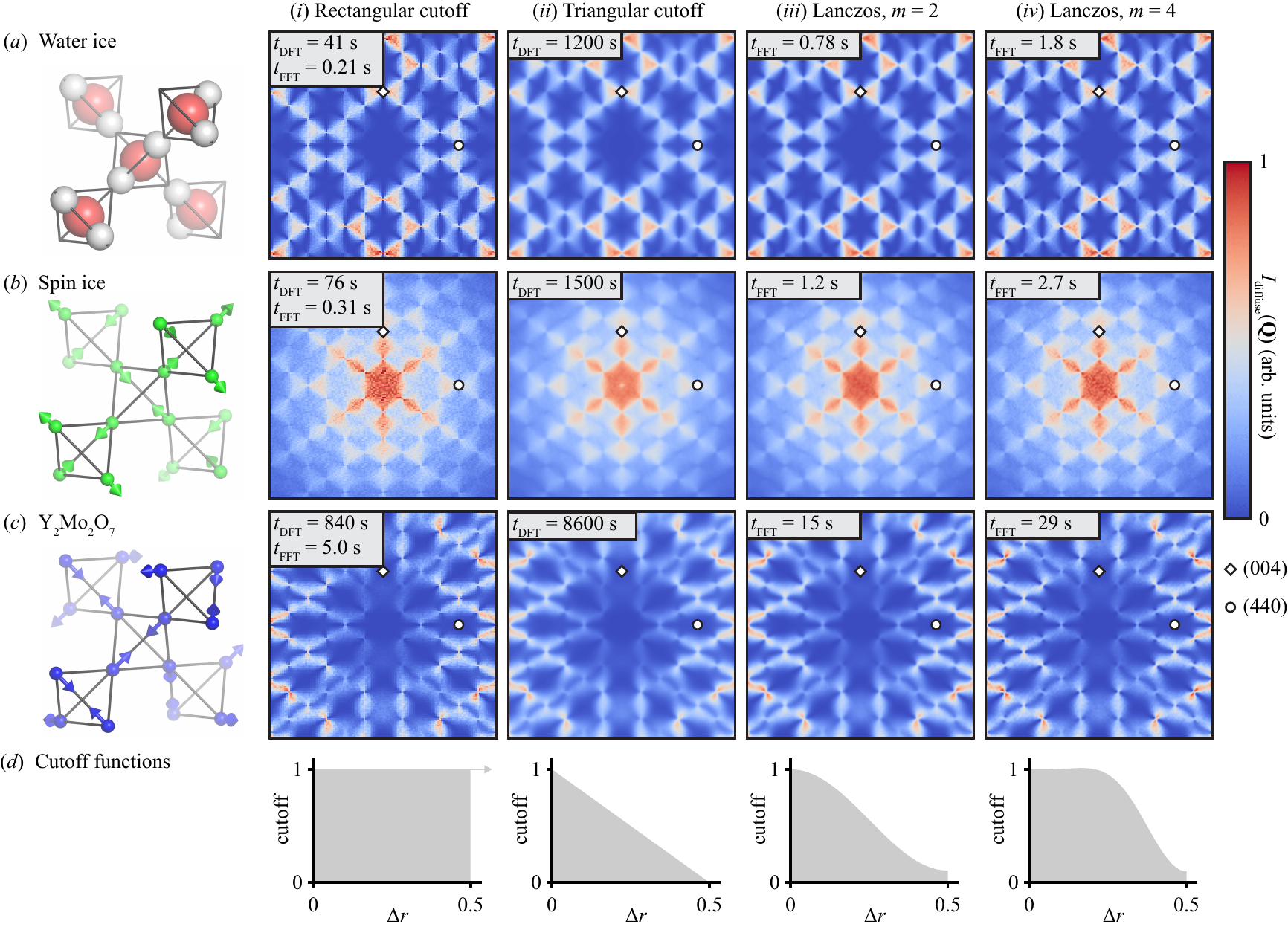}
\caption{\label{fig:fig1}Diffuse-scattering calculations and real-space cutoff functions for
different disorder models and methods of calculation. Rows (\emph{a})\textendash (\emph{c})
show results for three disorder models based on ``ice rules'': (\emph{a})
water ice (occupational disorder), (\emph{b}) spin ice (magnetic disorder),
and (\emph{c}) Y$_{2}$Mo$_{2}$O$_{7}$ (displacive disorder). Row
(\emph{d}) shows the cutoff function defined in Eq.~(\ref{eq:pdf_cutoff}),
where $\Delta r$ is the separation between atom pairs parallel to
a crystal axis. For each row, columns show the calculated diffuse-scattering
intensity $I_{\mathrm{diffuse}}(\mathbf{Q})$ obtained using different
methods, and the corresponding cutoff function. Column (\emph{i})
shows calculated intensities at the supercell Bragg positions obtained
using Eqs.~(\ref{eq:comp_diffuse_intensity}), (\ref{eq:mag_intensity}),
and (\ref{eq:disp_diffuse_intensity}) for rows (\emph{a})\textendash (\emph{c}),
respectively. The calculated intensities shown in (a)\emph{i}\textendash (c)\emph{i}
contain noise and are not smooth functions of $\mathbf{Q}$, because
the flat cutoff function shown in (\emph{d})\emph{i} does not suppress
long-range correlations that add to the noise. Column (\emph{ii})
shows results for the method of sub-boxes \cite{Butler_1992} discussed
in section \ref{subsec:subboxes}. The calculated intensities in (\emph{a})\emph{ii}\textendash (\emph{c})\emph{ii}
have low noise and are smooth functions of $\mathbf{Q}$, because
the triangular cutoff function shown in (\emph{d})\emph{ii} suppresses
long-range correlations. Columns (\emph{iii}) and (\emph{iv}) show
results for the Lanczos resampling approach discussed in section \ref{subsec:lanczos}
with $m=2$ and $m=4$, respectively, where $m$ is defined in Eq.~(\ref{eq:lanczos}). The Lanczos approach is an alternative to the
method of sub-boxes. The calculated intensities in (\emph{a})\emph{iii}\textendash (\emph{c})\emph{iii}
and (\emph{a})\emph{iv}\textendash (\emph{c})\emph{iv} again have
reduced noise and are smooth functions of $\mathbf{Q}$, because the
Lanczos cutoff functions shown in (\emph{d})\emph{iii} and (\emph{d})\emph{iv}
suppresses long-range correlations by an amount determined by the
parameter $m$. For each calculation of the diffuse-scattering intensity,
the CPU time required for a calculation using the discrete Fourier
transform $t_{\mathrm{DFT}}$ and/or the fast Fourier transform $t_{\mathrm{FFT}}$
is labelled. The FFT (+ optional Lanczos resampling) method produces
accurate results and reduces the CPU time by between two and three
orders of magnitude compared to the discrete Fourier transform (+
optional sub-boxes) method.}
\end{figure*}

Figure \ref{fig:fig1}(\emph{a})\emph{i}\textendash (\emph{c})\emph{i} shows water-ice,
spin-ice, and orbital-ice scattering calculations, respectively. The
results for water ice and spin ice are in good agreement with published
calculations \cite{Wehinger_2014,Fennell_2009}. Calculations using
the discrete Fourier transform and the FFT gave identical results,
as required. Crucially, however, the FFT-based calculations are much faster;
e.g., the CPU time required 
for the FFT-based calculation of the water-ice scattering was $0.21$\,s compared to $41$\,s for the discrete-Fourier-transform calculation\textemdash an
improvement of two orders of magnitude. Moreover, although 55 FFTs were required for the fifth-order approximation
in the orbital-ice calculation, a reduction in CPU time of two orders
of magnitude was still obtained. The largest error in any pixel for
this fifth-order approximation was less than $0.7\%$, confirming
that the Taylor expansion can be used for quantitative studies \cite{Butler_1993}.

To conclude this section, I discuss the performance of the FFT-based approach for scattering volumes and supercells that are much larger than those considered above.
While Figure \ref{fig:fig1} shows a scattering plane, large scattering
volumes can be calculated in a few minutes using the FFT; e.g., for the water-ice
model, a cube in reciprocal space extending from $-12\leq h,k,l\leq12$
and containing $\sim$$7\times10^{6}$ supercell Bragg positions was
calculated in $58$\,s. Moreover, the $n_{1}n_{2}n_{3}\log_{2}(n_{1}n_{2}n_{3})$ scaling of the FFT means that larger supercells yield greater improvements in speed compared to the discrete Fourier transform; e.g., an FFT-based calculation for the spin-ice model using $40\times40\times40$ supercells, but otherwise equivalent to that shown in figure \ref{fig:fig1}(\emph{b})\emph{i}, was faster than the corresponding discrete-Fourier-transform calculation by a factor of approximately $1400$.

\section{Noise reduction\label{sec:resampling}}

\subsection{Method of sub-boxes\label{subsec:subboxes}}

The scattering patterns shown in figure \ref{fig:fig1}(\emph{a})\emph{i}\textendash (c)\emph{i},
while of good quality, are affected by two problems. First, because
the diffuse scattering has been calculated only at supercell Bragg
positions, it is not a smooth function of wavevector. Second, the
calculations are visibly affected by high-frequency noise (``speckle''),
even though the results shown have been averaged over 100 independent
supercells to reduce noise. As discussed in section \ref{sec:introduction},
this noise occurs because the calculation can be dominated by the
many essentially-uncorrelated atom pairs that are large distances
apart. 

A proven approach to address these problems is to divide the supercell
into a set of smaller overlapping regions, called ``sub-boxes''
or ``lots'' \cite{Butler_1992,Proffen_1997}. Each sub-box has open
boundary conditions, and contains $n_{\mathrm{sb}}$ unit cells parallel
to each crystal axis ($n_{\mathrm{sb}}^{3}$ unit cells in total).
The value of $n_{\mathrm{sb}}$ should exceed the correlation length
of the local modulations that generate the diffuse scattering, but
be smaller than or equal to $n_{\alpha}/2$, to avoid contributions
from periodic images of the supercell. The intensity for each sub-box
is calculated separately using the discrete Fourier transform, and
the results averaged to estimate $I(\mathbf{Q)}$. In the original
algorithm, sub-boxes were distributed at random within the supercell,
and the number of sub-boxes chosen so that each atom was included
at least once \cite{Butler_1992}. However, this random-sampling approach
has the disadvantage that a different answer is obtained each time
the calculation is run. A recent algorithm addressed this problem
by using every lattice point in the supercell as the origin of a sub-box,
while avoiding unnecessary repetition of phase-factor calculations
\cite{Paddison_2013b}. I used this algorithm to perform scattering
calculations for the example models discussed in section \ref{sec:examples},
using a sub-box length $n_{\mathrm{sb}}=5$ unit cells.

Figure \ref{fig:fig1}(\emph{a})\emph{ii}\textendash (\emph{c})\emph{ii} shows example scattering calculations using the method of sub-boxes.
The use of open boundary conditions for each sub-box means that the
calculation is not restricted to the Bragg positions of the supercell.
Moreover, the calculation is essentially free of visible noise. There
are two potential reasons for this. First, it is possible in general
to choose $n_{\mathrm{sb}}<n_{\alpha}/2$, which excludes longer-range
correlations from the calculation \cite{Butler_1992}. However, since
I take $n_{\mathrm{sb}}=n_{\alpha}/2$, this reason does not apply
in the present example. In fact, the method of sub-boxes is effective
here because each sub-box has open boundary conditions, as I now discuss. I first recall that, for a periodic supercell, the number of distinct
pairs of lattice points separated by $n_{\alpha}\Delta r_{\alpha}$
unit cells (parallel to crystal axis $\alpha$) is equal to $n_{\alpha}$;
the cutoff function in Eq.~(\ref{eq:pdf_cutoff}) is therefore
a rectangular function, as shown in figure \ref{fig:fig1}(\emph{d})\emph{i}. In
contrast, if the supercell has open boundary conditions, then the
number of distinct pairs of lattice points separated by $n_{\alpha}\Delta r_{\alpha}$
unit cells is equal to $n_{\alpha}(1-\Delta r_{\alpha})$. Consequently,
the cutoff function for an open supercell is a triangular function
that is equal to unity at $\Delta r_{\alpha}=0$ and zero at $\Delta r_{\alpha}=\pm1$.
Hence, in the method of sub-boxes, the cutoff function is also triangular
and decays to zero at $\Delta r_{\alpha}=\pm n_{\mathrm{sb}}/n_{\alpha}$,
as shown in figure \ref{fig:fig1}(\emph{d})\emph{ii}. This is the main result of
this section. It explains how the method of sub-boxes reduces noise,
because as the separation of atom pairs increases, their pair correlations
are suppressed in the scattering calculation.

The analysis above also reveals two limitations of the method of sub-boxes.
First, it is very much slower than the FFT-based approach. Indeed, it is also much
slower than the discrete Fourier transform used to calculate $I(\mathbf{G})$
(figure \ref{fig:fig1}(\emph{a})\emph{i}\textendash (\emph{c})\emph{i}): this is
mainly because the $401\times401$ grid of wavevectors samples reciprocal
space at finer intervals than the Nyquist rate defined by the spacing
of the supercell Bragg positions, and hence calculates redundant information.
Second, its effectiveness at reducing noise has the trade-off that
the scattering is artificially blurred. Although this blurring can
be reduced by increasing $n_{\mathrm{sb}}$, it is never entirely
absent, because the triangular cutoff function is never flat. This
limitation may be significant if accurate estimates of the correlation
length are required. 

\subsection{Lanczos resampling\label{subsec:lanczos}}

I now develop an approach that addresses the limitations of the method
of sub-boxes, while maintaining its noise-reduction properties. Whereas
the method of sub-boxes works directly in real space, I will show
that its effects can be replicated efficiently by applying a resampling
filter in reciprocal space. 

I first recall that $I(\mathbf{Q})$ can, in principle, be calculated
at any wavevector $\mathbf{Q}$ given only its samples at the supercell
Bragg positions $\{\mathbf{G}\}$, by applying the Whittaker-Shannon
interpolation formulae, Eqs.~(\ref{eq:intensity_convol}) and
(\ref{eq:weight_sinc}). Consequently, $I(\mathbf{G})$ can be calculated
using the FFT and then resampled to obtain $I(\mathbf{Q})$ on a fine
grid. An apparent problem with this resampling approach is that Eq.~(\ref{eq:intensity_convol}) requires summation over an infinite number
of supercell Bragg positions to reproduce the discontinuity in the
rectangular cutoff function at $\Delta r_{\alpha}=\pm r_{\mathrm{cut}}$.
If the infinite summation is replaced by a finite summation, then
large truncation artifacts (Fourier ripples) are introduced in the
cutoff function, which do not vanish even as the number of summed
terms becomes very large; this undesirable effect is known as the
Gibbs phenomenon \cite{Gibbs_1899}. However, the key result of section
\ref{subsec:subboxes}\textemdash that the method of sub-boxes effectively applies a triangular cutoff function\textemdash suggests that
a rectangular cutoff function is not necessary or even desirable for
diffuse-scattering calculations. The triangular cutoff function, and
other suitably-chosen continuous functions, can be approximated by
using a finite summation in Eq.~(\ref{eq:intensity_convol}).
Consequently, the scattering intensity can be approximated as
\begin{align}
I(\mathbf{Q}) & \approx\frac{\sum_{\mathbf{G}}I(\mathbf{G})W(\mathbf{Q}-\mathbf{G})}{\sum_{\mathbf{G}}W(\mathbf{Q}-\mathbf{G})},\label{eq:intensity_convol_finite}
\end{align}
where the weight function $W(\mathbf{Q}-\mathbf{G})$ is the Fourier
transform of the cutoff function, and the summation runs over supercell
Bragg positions in the vicinity of $\mathbf{Q}$. This is the basis
for the technique called Lanczos resampling \cite{Lanczos_1956,Duchon_1979},
which I discuss below. 

The effects of Lanczos resampling in real and reciprocal space can
be derived in three steps \cite{Duchon_1979}. First, the desired
form of the cutoff function is specified. Second, $W(\mathbf{Q}-\mathbf{G})$
is determined as the Fourier transform of this idealized cutoff function.
Third, $W(\mathbf{Q}-\mathbf{G})$ is back-Fourier-transformed numerically
to determine the actual cutoff function obtained for a finite summation
range. Following Lanczos \cite{Lanczos_1956}, I consider an idealized
cutoff function that is equal to unity for $|\Delta r_{\alpha}|<0.5(1-2/m)$
and decays linearly to zero over the range $0.5(1-2/m)\leq|\Delta r_{\alpha}|\leq0.5$,
where $m\ge2$ is an integer \cite{Duchon_1979}. Hereafter, the nominal
cutoff separation $r_{\mathrm{cut}}=0.5(1-1/m)$ is the separation
at which the idealized cutoff function equals $0.5$. Since this cutoff
function is the convolution of two rectangular functions, its Fourier
transform is the product of two sinc functions, so that
\begin{equation}
W(\mathbf{Q}-\mathbf{G})=W_{\mathrm{sinc}}(\mathbf{Q}-\mathbf{G})\thinspace L(\mathbf{Q}-\mathbf{G}),\label{eq:windowed_sinc}
\end{equation}
where $W_{\mathrm{sinc}}(\mathbf{Q}-\mathbf{G})$ is given by Eq.~(\ref{eq:weight_sinc}) with $r_{\mathrm{cut}}=0.5(1-1/m)$,
and
\begin{align}
L(\mathbf{Q}-\mathbf{G}) & =\begin{cases}
\prod_{\alpha}\textrm{sinc}\left[\pi(Q_{\alpha}-G_{\alpha})/m\right] & \textrm{if }|Q_{\alpha}-G_{\alpha}|<m,\\
0 & \textrm{otherwise}
\end{cases}\label{eq:lanczos}
\end{align}
is called the window function. Because Eq.~(\ref{eq:lanczos})
is set to zero outside the central lobe of its sinc function \cite{Duchon_1979},
the summation in Eq.~(\ref{eq:intensity_convol_finite}) is only
taken over the $(2m)^{3}$ values of $\mathbf{G}$ for which $[Q_{\alpha}]-m+1\leq G_{\alpha}\leq[Q_{\alpha}]+m$,
where square brackets denote the floor function. This approach is
called windowed-sinc filtering. It has been widely used for image
filtering and digital-signal processing \cite{Getreuer_2011,Marschner_1994}.
It was applied to diffuse-scattering calculations for the first time
by the present author \cite{Paddison_2018}; a more detailed analysis
is given here.

Example scattering calculations using Lanczos resampling with $m=2$
are shown in figure \ref{fig:fig1}(\emph{a})\emph{iii}\textendash (\emph{c})\emph{iii}
and calculations with $m=4$ are shown in figure \ref{fig:fig1}(\emph{a})\emph{iv}\textendash (\emph{c})\emph{iv}.
In each case, $I(\mathbf{G})$ was calculated using the FFT, and $I(\mathbf{Q})$
was then estimated using Eqs.~(\ref{eq:weight_sinc}) and (\ref{eq:intensity_convol_finite})\textendash (\ref{eq:lanczos}).
The corresponding cutoff functions are shown in figure \ref{fig:fig1}(\emph{d})\emph{iii}
and 1(\emph{d})\emph{iv}. The idealized cutoff function for the $m=2$
case is a triangular function, and the actual cutoff function resembles
a smoothed version of it. Consequently, the scattering calculated
by Lanczos resampling with $m=2$ closely resembles the sub-box calculation.
The Lanczos approach is, however, faster than the method of sub-boxes
by approximately three orders of magnitude. This large improvement
arises first from the use of the FFT, and second because the scattering
is only calculated at the supercell Bragg positions before being resampled
to the $401\times401$ pixel grid. As the value of $m$ is increased,
the cutoff function increasingly resembles the rectangular function;
hence, by tuning the value of $m$, it is possible to trade off the
level of noise that can be tolerated against the amount of blur that
is introduced. In particular, choosing $m>2$ ensures that the cutoff
function is essentially flat for small $\Delta r_{\alpha}$, so that
correlations at near-neighbour distances can be accurately reflected
in the scattering pattern.

I mention three possible modifications to the Lanczos resampling approach.
First, if $I(\mathbf{Q})$ is calculated on a grid with axes parallel
to the reciprocal-lattice vectors, Eqs.~(\ref{eq:windowed_sinc})
and (\ref{eq:lanczos}) can be calculated along each axis in sequence,
which reduces the computational expense associated with resampling
\cite{Getreuer_2011}. Second, whereas the equations above assume
that the idealized cutoff function decays to zero at $\Delta r_{\alpha}=0.5$,
it may be preferable to choose a cutoff function that decays to zero
at $\Delta r_{\alpha}<0.5$ if further noise suppression is required.
This corresponds to choosing $n_{\mathrm{sb}}<n_{\alpha}/2$ in the
method of sub-boxes. In the Lanczos approach, an idealized cutoff
function with $r_{\mathrm{cut}}=(1-1/m)/m^{\prime}$ that decays
to zero at $|\Delta r_{\alpha}|=1/m^{\prime}$ can be obtained for
integer $m^{\prime}>2$ by replacing $m$ with $mm^{\prime}/2$ in
Eq.~(\ref{eq:lanczos}). Finally, it may sometimes be preferable
to use a spherically-symmetric cutoff function. This can be achieved
by replacing Eq.~(\ref{eq:weight_sinc}) with 
\begin{equation}
W_{\mathrm{r}}(|\mathbf{Q}-\mathbf{G}|)=\frac{4\pi|\mathbf{r}_{\mathrm{c}}|^{3}}{n_{1}n_{2}n_{3}V}\left[\frac{\sin(|\mathbf{Q}||\mathbf{r}_{\mathrm{c}}|)}{(|\mathbf{Q}||\mathbf{r}_{\mathrm{c}}|)^{3}}-\frac{\cos(|\mathbf{Q}||\mathbf{r}_{\mathrm{c}}|)}{(|\mathbf{Q}||\mathbf{r}_{\mathrm{c}}|)^{2}}\right],
\end{equation}
where $|\mathbf{r}_{\mathrm{c}}|=(1-1/m)|\mathbf{r}_{\mathrm{max}}|$,
$|\mathbf{r}_{\mathrm{max}}|$ is the radius of the largest sphere
that can be inscribed in the supercell, and $n_{1}n_{2}n_{3}V$ is
the supercell volume \cite{Marschner_1994}. 

\section{Conclusions\label{sec:conclusions}}

The main result of this article is that diffuse-scattering calculations
from atomistic models can be accelerated by at least two orders of
magnitude by applying two fundamental properties of scattering patterns.
The first is that the average lattice of a crystal has the periodicity
of the crystallographic unit cell, even in the presence of disorder;
this property allows the fast Fourier transform (FFT) algorithm to
be used to calculate diffuse scattering, which has not previously
been considered feasible \cite{Butler_1992}. The second is that the
scattering intensity of a periodic supercell is determined at all
wavevectors by its values at the Bragg positions of the supercell;
this observation allows the scattering intensity to be determined
on an arbitrarily fine wavevector grid without redundant calculations.
These observations are by no means new: indeed, they date to the earliest
work on structural disorder \cite{Borie_1971} and sampling theory
in crystallography \cite{Sayre_1952}. Nevertheless, their importance
for diffuse-scattering calculations has not previously been recognized.

The acceleration of diffuse-scattering calculations enabled by these
results and the associated computer program \textsc{Scatty} [27]
has important practical implications. Comparisons of model calculations
with the large 3D datasets obtained by modern X-ray and neutron instruments
would previously have required many hours, but can now be performed
in a few minutes on a laptop. Moreover, it is now practical to fit
models of the interactions between disorder variables (local displacements,
occupancies, or magnetic moments) directly to 3D experimental datasets
in a matter of hours; such fits would previously have required many
weeks or months, which has proved prohibitive \cite{Welberry_1998,Welberry_2008}.
It is also possible to parallelize the intensity calculations for
many supercells, potentially leading to further reductions in computer
time \cite{Gutmann_2010}. I therefore anticipate that these developments
will make quantitative analysis of large diffuse-scattering datasets
more practical, more quantitative, and more easily automated. I hope
that, similar to the advent of 3D data analysis in conventional crystallography
approximately 60 years ago, the prospect of 3D diffuse-scattering analysis identified previously \cite{Welberry_2008}
will allow the diffuse-scattering approach to be applied to a much wider range of disordered
crystalline materials than has yet been possible.

\acknowledgements{I would like to thank Andrew Goodwin, Ross Stewart, Matthew Cliffe, and Arkadiy Simonov for valuable
discussions related to this work.}

\end{document}